\documentclass[aps,onecolumn,showpacs,preprintnumbers,amsmath,amssymb,superscriptaddress,floatfix,nofootinbib]{revtex4}

\usepackage{graphicx}
\usepackage{amsmath}
\usepackage{amsfonts}
\usepackage{amssymb}
\usepackage{color}

\begin{document}

\title{Role of the hidden charm $N^*_{c\bar{c}}(4261)$ resonance  in the $\pi^- p \to J/\psi n$ reaction}

\author{Zhen Ouyang}~\email{ouyangzh@impcas.ac.cn}
\affiliation{Institute of Modern Physics, Chinese Academy of
Sciences, Lanzhou 730000, China}
%\affiliation{Research Center for
%Hadron and CSR Physics, Institute of Modern Physics of CAS and
%Lanzhou University, Lanzhou 730000, China} \affiliation{State Key
%Laboratory of Theoretical Physics, Institute of Theoretical Physics,
%Chinese Academy of Sciences, Beijing 100190, China}
\author{Li-Ping Zou}
\affiliation{Institute of Modern Physics, Chinese Academy of
Sciences, Lanzhou 730000, China}
\affiliation{Institute of Basic Sciences, Konkuk University,
Seoul 151-747, Korea}

\begin{abstract}

We employ an effective Lagrangian approach and isobar model to investigate the role
of the $N^*_{c\bar{c}}(4261)$ resonance with hidden charm in the $\pi^- p \to J/\psi n$ reaction.
The total and differential cross sections of this reaction are predicted
by including contributions from both $N^*_{c\bar{c}}(4261)$ and nucleon pole.
It is found that the maximal value of the total cross section can exceed 11 $\mu b$
and a clear $N^*_{c\bar{c}}(4261)$ peak is visible there, well distinguished from background.
As center-of-mass energy increases to about 4.8 GeV, the background makes the differential cross section for backward angles 
rather different. Our theoretical results would provide valuable information for
looking for the $N^*_{c\bar{c}}(4261)$ resonance in future experiments.

\end{abstract}

\date{\today}

%\pacs{12.40.Vv}

\maketitle

\section{Introduction}
In the conventional quark model, all established baryons are regarded as three-quark ($qqq$) configurations. However, in principle, the fundamental theory of the strong interaction QCD indeed allows existence of the non $q\bar{q}$ and $qqq$ states which are beyond the quark model and called exotic states. In fact, some scientists have already suggested certain baryon resonances to be meson-baryon dynamically generated states~\cite{Weise,or,Oset, meiss,Inoue, lutz,Hyodopk} or states with large pentaquark ($qqqq\bar q$) components~\cite{Riska,Liubc,Zou10}, which could cure various problems and difficulties that people meet when they applied the constituent quark model to explanations of experiments. Thus, it is very necessary to distinguish these proposed exotic baryons from those three-quark ($qqq$) states predicted in diverse quenched quark models. The case is more complicated in the light flavor sector. Since all the baryon resonances generated in various models mentioned above are composed of light quarks, no matter three or five constituent quarks, hence they have masses around the same energy region. In addition, all these models always have some tunable ingredients to fit the experimental data. As a consequence, it is hard to discern the correct ones from so many model predictions and nail down the structure of these baryon resonances. The situation will be very different for the hidden charm baryons with the figures $qqqc\bar c$.
Recently, the authors of Refs.~\cite{Wu:2010vk,Wu:2010jy} predicted a series of hidden charm $N^*$ and $\Lambda^*$ resonances with masses above 4.2 GeV and widths smaller than 100 MeV, which are dynamically generated by using the dynamics of the local hidden gauge Lagrangians in combination with the coupled-channel unitary approach. In this paper we will concentrate on one of them, namely the $N^*_{c\bar{c}}(4261)$ resonance, which was also predicted under the consideration of heavy quark spin symmetry in Ref.~\cite{Xiao:2013yca}. Since the $N^*_{c\bar{c}}(4261)$ resonance with hidden charm
has a much larger mass than normal $N^*$ resonances, which implies large $c\bar{c}$ components in the $N^*_{c\bar{c}}(4261)$, and a relatively small width, it definitely cannot be accomodated by the conventional quark models.

In view of the intriguing and distinctive features of the $N^*_{c\bar{c}}(4261)$ resonance, it is meaningful for us to ask the question where we can search for it. In Refs.~\cite{Wu:2010vk,Wu:2010jy} where Wu {\sl et al.} presented the first prediction of the $N^*_{c\bar{c}}(4261)$, they have simultaneously proposed to produce this hidden charm baryon with proton-antiproton collisions in the scheduled experiments at PANDA/FAIR. After estimating the production cross sections of the $N^*_{c\bar{c}}(4261)$ resonance in the $\bar{p}p \to \bar{p}p \eta_c$ and $\bar{p}p \to \bar{p}p J/\psi$ reactions, they concluded that it could be observed by the PANDA/FAIR. Huang {\sl et al.}~\cite{Huang:2013mua} studied the discovery potential of this $N^*_{c\bar{c}}(4261)$ resonance at JLab 12 GeV and after detailed
calculations, unfortunately, they found that it is difficult to look for $N^*_{c\bar{c}}(4261)$ in photoproduction due to extremely large background.

In the near future, the pion-proton scattering experiments will be carried out at J-PARC which is expected to provide pion beam of energy over 20 GeV~\cite{Kim:2014qha}. Therefore, the J-PARC in Japan is sufficient to generate a variety of hidden charm baryons and will surely offer another powerful tool besides PANDA/FAIR to search for the $N^*_{c\bar{c}}(4261)$ resonance. On the theoretical side, the authors of Ref.~\cite{Wang:2015qia} have explored the production of $N^*_{c\bar{c}}(4261)$ in the $\pi^- p \to \eta_c n$ process, and they found that a clear peak, easily distinguished from background, is shown in the total cross section at the center-of-mass energy $W \approx 4.3$ GeV. Moreover, in Ref.~\cite{Garzon}, the role of the $N^*_{c\bar{c}}(4261)$
resonance was investigated in the $\pi^-
p \to D^- \Sigma^+_c$ reaction, and it was found that the $N^*_{c\bar{c}}(4261)$ resonance results in a significant enhancement of the total cross section
near threshold.

Inspired by the above two theoretical works~\cite{Wang:2015qia,Garzon}, since the $N^*_{c\bar{c}}(4261)$ can also decay into the open channel $J/\psi N$ according to Refs.~\cite{Wu:2010vk,Wu:2010jy,Xiao:2013yca}, here we study the role and
production of the hidden charm $N^*_{c\bar{c}}(4261)$ resonance in the $\pi^- p \to J/\psi n$ reaction, based on an effective Lagrangian approach and isobar model.

The article is organized as follows. After the introduction, we give the
formalism and ingredients used in our calculations. The numerical results
and discussion are presented in Section III. Finally, the article ends with
some conclusions.

\section{Formalism and ingredients}

\begin{figure}[htbp]
\begin{center}
\includegraphics*[scale=0.7]{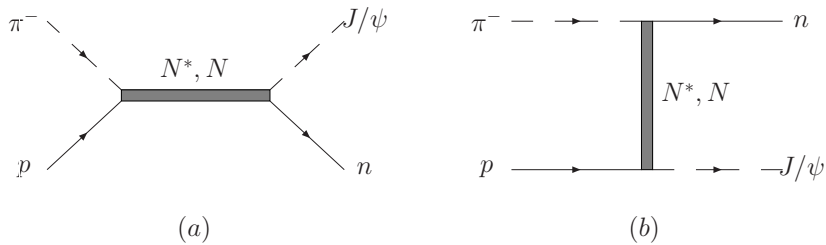}
\caption{Feynman diagrams considered in the $\pi^- p \to J/\psi n$ reaction. }
\label{fig:diagrams}
\end{center}
\end{figure}

In this work, we employ the effective Lagrangian approach together with the isobar model to study
the $\pi^- p \to J/\psi n$ process. All the Feynman
diagrams of tree level involved for this process are depicted in Fig.~\ref{fig:diagrams}.
The hidden charm $N^*_{c\bar{c}}(4261)$ ($\equiv N^*$) resonances are produced as intermediate
states via $s$ and $u$ channels as shown in Fig.~\ref{fig:diagrams}(a) and Fig.~\ref{fig:diagrams}(b), respectively.
Besides, we have to add the main background contributions for the creation of $N^*_{c\bar{c}}(4261)$ in our calculations. For this purpose,
the production of the intermediate virtual nucleon is also considered with the same mechanism as
the production of $N^*_{c\bar{c}}(4261)$.

In the theoretical frame of Refs.~\cite{Wu:2010vk,Wu:2010jy,Xiao:2013yca},
only S-wave interactions of mesons and baryons are considered to dynamically generate the $N^*_{c\bar{c}}(4261)$, so its spin-parity $J^P$ is set
to be $1/2^-$. To evaluate the amplitudes of diagrams in Fig.~\ref{fig:diagrams} within the isobar model, we need to
know the effective Lagrangians involving the $N^*_{c\bar{c}}(4261)$ resonance. Here we apply a Lorentz covariant orbital-spin (L-S) scheme for
$N^* NM$ couplings to get the effective $N^*_{c\bar{c}}(4261)N \pi$ and $N^*_{c\bar{c}}(4261)N J/\psi$ vertices appearing in
the Feynman diagrams depicted in Fig.~\ref{fig:diagrams}. In this covariant L-S scheme~\cite{zouprc03}, the effective $N^* NM$ Lagrangians were constructed for
excited nucleon states $N^*$ of arbitrary spin-parity (such as $1/2^-$) decaying
into the meson-nucleon final states with a definite orbital angular momentum, and the pure S-wave $N^* NM$ couplings given there are just suitable for the case of $N^*_{c\bar{c}}(4261)$. These S-wave couplings have already been used to explore the photoproduction of
$N^*_{c\bar{c}}(4261)$ and other hidden charm baryons in Ref.~\cite{Huang:2013mua}. The explicit forms of the effective $N^*_{c\bar{c}}(4261)N \pi$ and $N^*_{c\bar{c}}(4261)N J/\psi$ couplings adopted in this work are taken from Ref.~\cite{Huang:2013mua} directly, given by
\begin{eqnarray}
{\mathcal L}_{N^*_{c\bar{c}}(4261)N \pi} &=& g_{N^*_{c\bar{c}}(4261)N \pi}
\overline{u}_{N} \vec\tau \cdot \vec\psi_{\pi} u_{N^*_{c\bar{c}}(4261)} + \text{h.c.},
\label{1440sig} \\
{\mathcal L}_{N^*_{c\bar{c}}(4261)N J/\psi} &=& ig_{N^*_{c\bar{c}}(4261)N J/\psi}
\overline{u}_{N} \gamma_5
\left(\gamma_{\mu}-\frac{q_{\mu} \not\!q}{q^2}\right) \psi^{\mu}_{J/\psi} u_{N^*_{c\bar{c}}(4261)} + \text{h.c.}, \label{1440pi}
\end{eqnarray}
where $u_N$, $u_{N^*_{c\bar{c}}(4261)}$, $\psi_{\pi}$, and
$\psi^{\mu}_{J/\psi}$ denote the  wave functions for the nucleon, $N^*_{c\bar{c}}(4261)$
resonance, $\pi$ and $J/\psi$-meson,
respectively, $q$ is the four momentum of the $N^*_{c\bar{c}}(4261)$ resonance, $\vec\tau$ is the Pauli matrix.
We take the coupling constant $g_{N^*_{c\bar{c}}(4261)N \pi}= 0.103$ from Refs.~\cite{Wang:2015qia,Garzon}, which was determined by the predicted
partial decay width of the $N^*_{c\bar{c}}(4261)$ resonance to the $\pi N$ channel, $\Gamma \left(N^*_{c\bar{c}}(4261) \to N \pi \right)=3.8$ MeV given by Ref.~\cite{Wu:2010jy}. In addition, we can get the coupling constant $g_{N^*_{c\bar{c}}(4261)N J/\psi}$ from the formula for the
spin-averaged transition probability of $N^*_{c\bar{c}}(4261)$ decaying into a nucleon and a $J/\psi$:
\begin{equation}
\frac{1}{2}\sum_{spins}|{\mathcal M}\left( N^*_{c\bar{c}}(4261) \to N J/\psi \right)|^2=
\frac{g^2_{N^*_{c\bar{c}}(4261)N J/\psi}(E_N+m_N)((M_{N^*_{c\bar{c}}(4261)}-E_N)^2+2m^2_{J/\psi})}{2m_Nm^2_{J/\psi}},
\label{1440d}
\end{equation}
with
\begin{equation}
E_N=\frac{M^2_{N^*_{c\bar{c}}(4261)} + m^2_N - m^2_{J/\psi}}{2M_{N^*_{c\bar{c}}(4261)}}, \label{1232d}
\end{equation}
with $E_N$ as the energy of the nucleon in the rest frame
of the $N^*_{c\bar{c}}(4261)$ resonance. Ref.~\cite{Xiao:2013yca} has predicted the spin-averaged transition probability
$\frac{1}{2}\sum_{spins}|{\mathcal M}\left( N^*_{c\bar{c}}(4261) \to N J/\psi \right)|^2= |g_i\left( N^*_{c\bar{c}}(4261) \to N J/\psi \right)|^2$ with
$|g_i\left( N^*_{c\bar{c}}(4261) \to N J/\psi \right)|=0.76$. Thus we obtain $g_{N^*_{c\bar{c}}(4261)N J/\psi}=0.417$ with $M_{N^*_{c\bar{c}}(4261)}=4261.87$ MeV~\cite{Xiao:2013yca}.

The effective Lagrangian densities for describing the $NN \pi$ and $NN J/\psi$ vertices are introduced as
\begin{equation}
{\mathcal L}_{NN \pi}  = ig_{NN \pi} \overline{u}_{N}
\gamma_5 \vec\tau \cdot \vec\psi_{\pi}
u_N , \label{piNN}
\end{equation}
\begin{equation}
{\mathcal L}_{NN J/\psi}  = -g_{NN J/\psi}  \overline{u}_{N}\gamma_{\mu} \psi^{\mu}_{J/\psi} u_N, \label{sigNN}
\end{equation}
with $g_{NN \pi}=13.45$ as used in Refs.~\cite{Wang:2015qia,mach}. Besides, in this work we adopt the same value of the coupling constant
$g_{NN J/\psi}=1.62 \times 10^{-3}$ as in
Refs.~\cite{Barnes,linqy}, which was estimated by the
measured partial decay width of $J/\psi \to p\bar{p}$~\cite{pdg2014}.

To take the virtuality of the
nucleon pole and $N^*_{c\bar{c}}(4261)$ resonance into account, we introduce the same form factor for each intermediate
baryon as taken in Refs.~\cite{feusterprc58,feusterprc59,shkprc72}:
\begin{equation}
F_B(q^2)=\frac{\Lambda^4_B}{\Lambda^4_B+ (q^2-M_B^2)^2},
\end{equation}
with $q$ being the four momentum of the intermediate baryon, namely the nucleon or $N^*_{c\bar{c}}(4261)$ in $s$-channel or $u$-channel.
$M_B$ and $\Lambda_B$
are the mass and cut-off parameter for the intermediate baryon, respectively.
In our concrete calculations, we use all the cut-off
parameters $\Lambda_B = \Lambda_{N} = \Lambda_{N^*_{c\bar{c}}(4261) }= 2.5$ GeV for both $s$ and $u$ channels in order to minimize the free
parameters, as done in Ref.~\cite{Garzon}.

In our amplitude calculations of Feynman diagrams in
Fig.~\ref{fig:diagrams}, we adopt the propagators for the intermediate nucleon and $N^*_{c\bar{c}}(4261)$ resonance as the following:
\begin{equation}
G_{N^*_{c\bar{c}}(4261)}(q) =  \frac{\not\!q + M_{N^*_{c\bar{c}}(4261)}}{q^2 -
M^2_{N^*_{c\bar{c}}(4261)} + iM_{N^*_{c\bar{c}}(4261)} \Gamma_{N^*_{c\bar{c}}(4261)}},
\label{4261}
\end{equation}
\begin{equation}
G_N(q)=\frac{\not\!q + M_N}{q^2-M^2_N},
\label{938}
\end{equation}
with $M_{N^*_{c\bar{c}}(4261)}=4261.87$ MeV and $\Gamma_{N^*_{c\bar{c}}(4261)}=56.9$ MeV according to Refs.~\cite{Wu:2010vk,Wu:2010jy,Xiao:2013yca}.

After the effective Lagrangian densities, coupling constants, form factors and propagators fixed, the
amplitudes for various diagrams shown in Fig.~\ref{fig:diagrams} can be written down straightforwardly by following
the Feynman rules. We present the
explicit expression of each amplitude for the intermediate nucleon or $N^*_{c\bar{c}}(4261)$ ($\equiv N^*$) production via $s$-channel or $u$-channel
as the following:
\begin{eqnarray}
{\mathcal M}^{N^*}_s& &= \sqrt{2} g_{N^*N\pi}
g_{N^*NJ/\psi}F_{N^*}(q^2_s)
\varepsilon_{\mu}(p_{J/\psi}) \bar{u}(p_n) \gamma_5
\left(\gamma^{\mu}-\frac{q^{\mu}_s \not\!q_s}{q^2_s}\right) G_{N^*}(q_s)
u(p_p),\\
{\mathcal M}^{N^*}_u& &= \sqrt{2} g_{N^*N\pi}
g_{N^*NJ/\psi}F_{N^*}(q^2_u)
\varepsilon_{\mu}(p_{J/\psi}) \bar{u}(p_n) G_{N^*}(q_u) \gamma_5
\left(\gamma^{\mu}-\frac{q^{\mu}_u \not\!q_u}{q^2_u}\right)
u(p_p),\\
{\mathcal M}^{N}_s& &= \sqrt{2} g_{NN\pi}
g_{NNJ/\psi}F_{N}(q^2_s)
\varepsilon^{\mu}(p_{J/\psi}) \bar{u}(p_n)
\gamma_{\mu} G_{N}(q_s)\gamma_5
u(p_p),\\
{\mathcal M}^{N}_u& &= \sqrt{2} g_{NN\pi}
g_{NNJ/\psi}F_{N}(q^2_u)
\varepsilon^{\mu}(p_{J/\psi}) \bar{u}(p_n)\gamma_5 G_{N}(q_u)\gamma_{\mu}
u(p_p),
\end{eqnarray}
where $u(p_p)$, $u(p_n)$ and $\varepsilon^{\mu}(p_{J/\psi})$ stand for
the spin wave functions of the initial proton, outgoing neutron and $J/\psi$-meson in the final
state, respectively. Here the subscripts s and u refer to the s and u channels for the $\pi^- p \to J/\psi n$ process, respectively.
And the total amplitude for the $\pi^- p \to J/\psi n$ reaction is just the simple sum of the four amplitudes given above, namely $\mathcal M={\mathcal M}^{N^*}_s+{\mathcal M}^{N^*}_u+{\mathcal M}^{N}_s
+{\mathcal M}^{N}_u$.

Then the unpolarized differential cross section in the center of mass (${\rm c.m.}$)
frame for the $\pi^- p \to J/\psi n$ reaction is given by the expression
\begin{equation}
\frac{d \sigma}{d \text{cos} \theta}=
\frac{m_p m_n}{16 \pi s}
\frac{|\vec{p}\,^{\rm{c.m.}}_{J/\psi}|}{|\vec{p}\,^{\rm{c.m.}}_\pi|}
{\sum_{spins}}|\mathcal{M}|^2,
\end{equation}
where $s=q^2_s=W^2$ is the Lorentz-invariant Mandelstam variable, and $\theta$ represents the scattering angle
of the outgoing $J/\psi$-meson relative to the pion beam direction in the ${\rm c.m.}$ frame, while
$\vec{p}\,^{\rm{c.m.}}_\pi$ and $\vec{p}\,^{\rm{c.m.}}_{J/\psi}$ are the
three momenta of $\pi^-$ and $J/\psi$ in the ${\rm c.m.}$ frame.
\section{Numerical Results and discussion}
\begin{figure}[htbp]
\begin{center}
\includegraphics[scale=0.7]{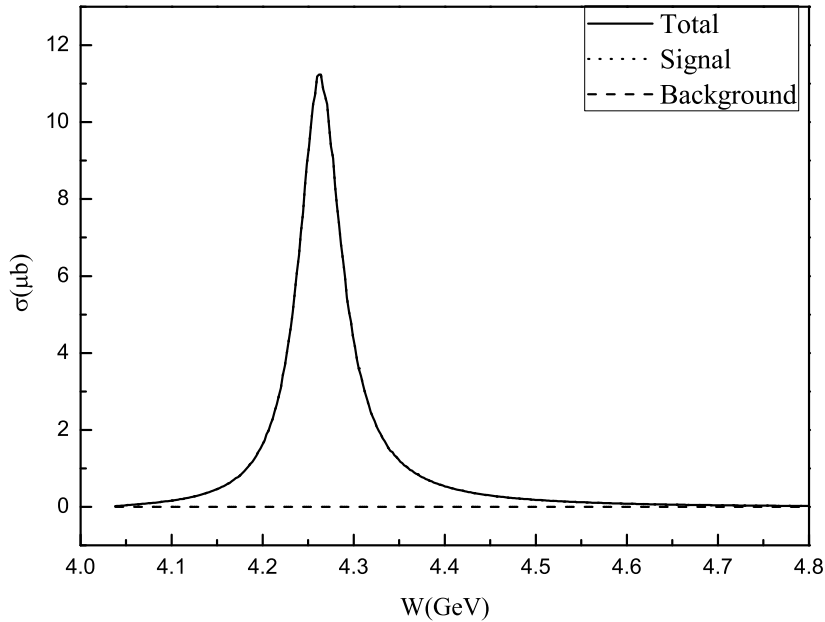}
\end{center}
\caption{Total cross section vs $W = \sqrt{s}$ for the $\pi^- p \to J/\psi n$ reaction compared with
the signal and background contributions. The solid, dotted and dashed lines stand for the total cross section, the signal contribution
from $N^*_{c\bar{c}}(4261)$ and the background contribution from nucleon pole, respectively.} \label{tcs1}
\end{figure}

\begin{figure}[htbp]
\begin{center}
\includegraphics[scale=0.7]{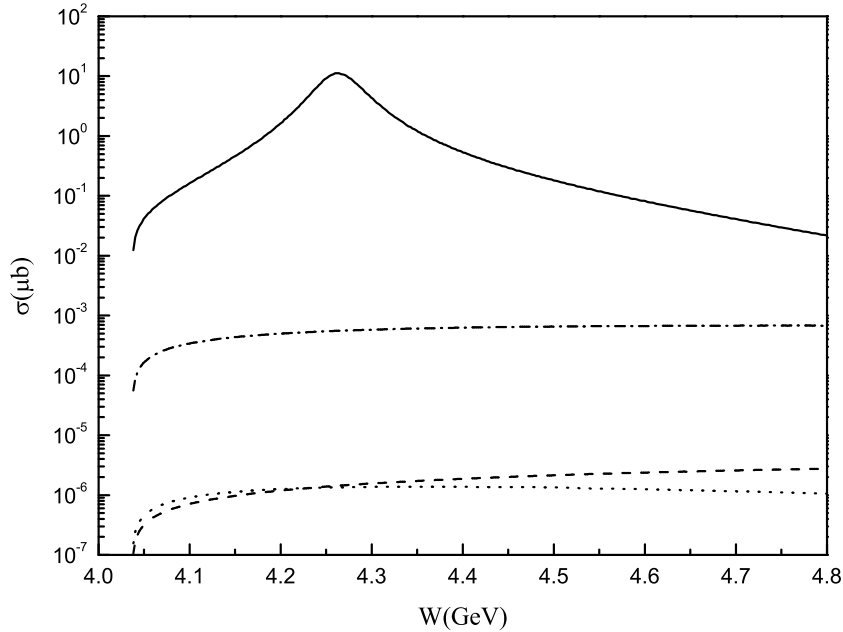}
\end{center}
\caption{Contributions of various components as a function of $W$ for the $\pi^- p \to J/\psi n$ reaction. The solid and dashed lines denote
contributions from $s$ and $u$-channel $N^*_{c\bar{c}}(4261)$ productions, respectively; while the dotted and dot-dashed curves represent
contributions from intermediate nucleon productions via $s$ and $u$ channels, respectively.} \label{tcs2}
\end{figure}

With the formalism and ingredients determined in the former section, we calculated the total
cross section versus the
invariant mass $W = \sqrt{s}$ of the $\pi^- p$ system for the $\pi^- p \to J/\psi n$ reaction.
Figs.~\ref{tcs1} and~\ref{tcs2} show our theoretical results for $W$ ranging from threshold to 4.8 GeV.

In Fig.~\ref{tcs1}, the total cross section including all the diagrams shown in Fig.~\ref{fig:diagrams} is presented along with the signal contribution
from $N^*_{c\bar{c}}(4261)$ production and the background contribution from intermediate nucleon production for
comparison. The numerical results for the total cross section, the signal and background contributions are
represented by solid, dotted and dashed curves, respectively. A peak due to $N^*_{c\bar{c}}(4261)$ production via $s$-channel is clearly visible in the
total cross section at center-of-mass energies around 4.3 GeV. And it is worth noting that the total cross section can reach up to approximately 11
$\mu b$ around 4.3 GeV. Moreover, it is found that the background makes several orders of magnitude smaller contribution to the total cross section than
the signal for $W$ between 4.1 and 4.4 GeV hence its contribution is negligible, which causes the coincidence of solid and dotted lines for
the total cross section and the signal in Fig.~\ref{tcs1}.

In Fig.~\ref{tcs2}, individual contributions from $N^*_{c\bar{c}}(4261)$ productions through
$s$ and $u$ channels are shown by solid and dashed curves, respectively; while contributions
corresponding to $s$ and $u$-channel productions of the nucleon pole are demonstrated by dotted and dot-dashed lines, respectively.
The contribution from the $N^*_{c\bar{c}}(4261)$ production via $s$-channel is found to be dominant over the whole energy region,
and there exists explicit cusp structure. In comparison, contribution from the $N^*_{c\bar{c}}(4261)$ production via $u$-channel
is extremely small. This is mainly because $q^2_u$ of the $N^*_{c\bar{c}}(4261)$ keeps negative for any $W = \sqrt{s}$, after subtracting
the square of large $N^*_{c\bar{c}}(4261)$ mass, which leads to strong suppression of its
propagator. The contribution from the intermediate nucleon production through $s$-channel is comparable to that from
$N^*_{c\bar{c}}(4261)$ production via $u$-channel and hence can be ignored, partly because this process is OZI suppressed, and
partly because the center-of-mass energy of the nucleon pole is above 4.0 GeV, far away from the nucleon mass-shell.
Furthermore, from Fig.~\ref{tcs2}, we can see that the $u$-channel production of the nucleon pole forms the most
important background for $\pi^- p \to J/\psi n$, although its contribution is still much smaller than the total cross section.

\begin{figure}[htbp]
\begin{center}
\includegraphics[scale=0.7]{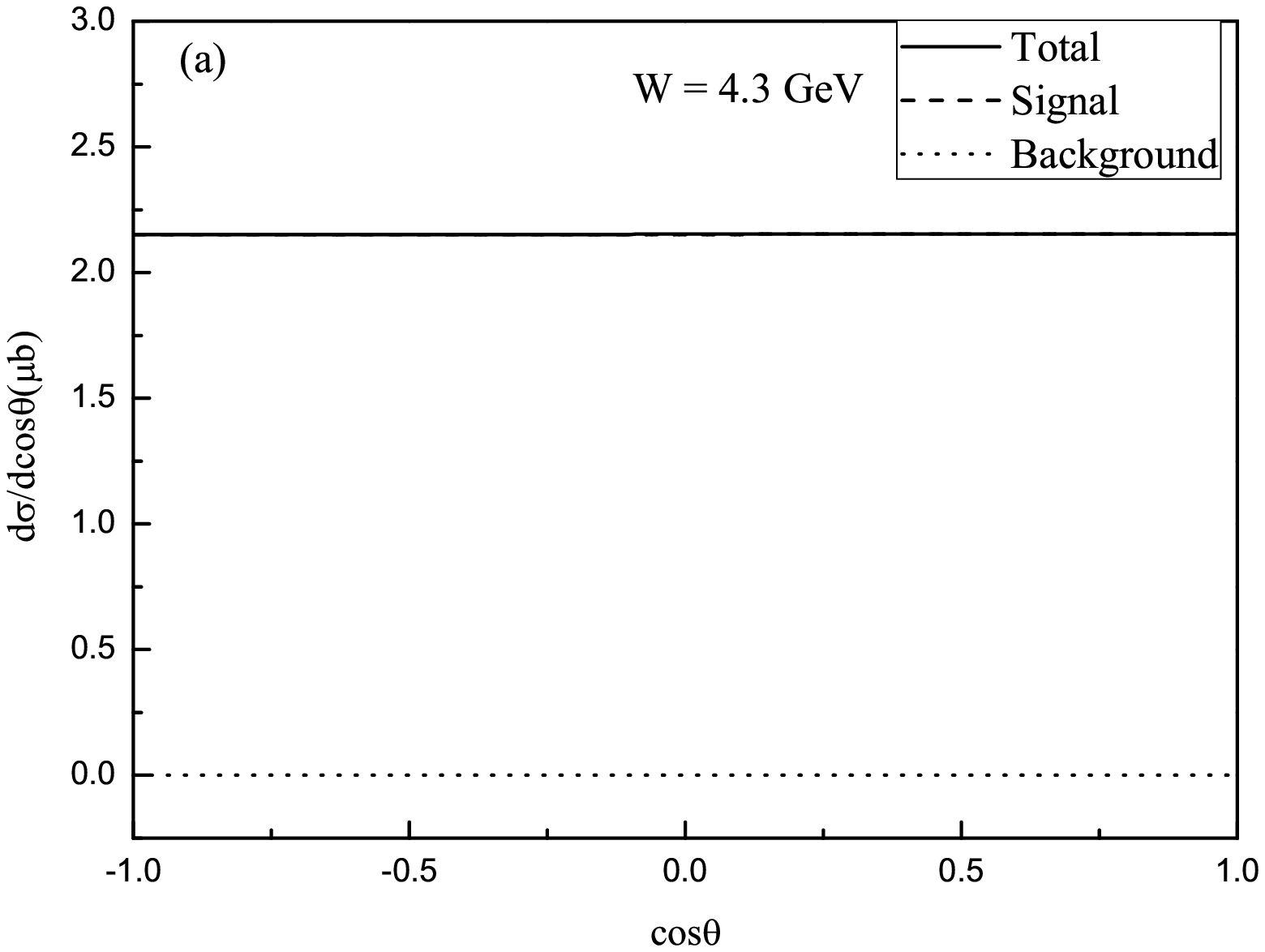}
\includegraphics[scale=0.7]{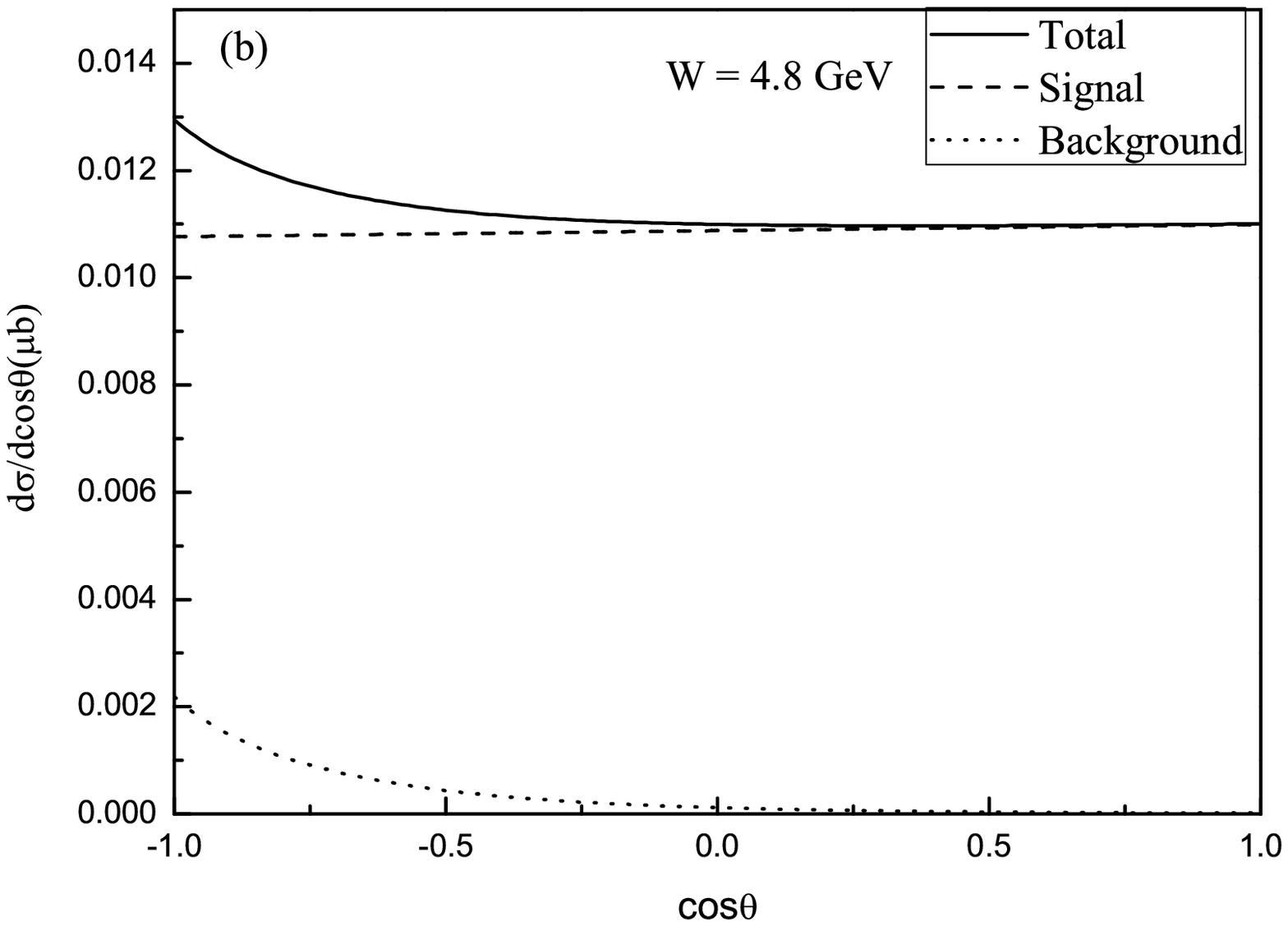}
\end{center}
\caption{Differential cross sections of $\pi^- p \to J/\psi n$
reaction at $W = 4.3$ GeV (a) and 4.8 GeV (b) in the ${\rm c.m.}$ frame, where "Total" represents
differential cross section including both signal and background contributions.}
\label{fig:dcs}
\end{figure}

In addition to the total cross section, we have also evaluated the differential cross section of this reaction
as a function of $cos \theta$ for two energy points.
Since the $N^*_{c\bar{c}}(4261)$ peak dominates the total cross section around 4.3 GeV, it is interesting to
further explore the corresponding differential cross section at this energy point. Then the contribution
from the $s$-channel $N^*_{c\bar{c}}(4261)$ production drops steeply as energy increases and the background contribution
becomes important around 4.8 GeV. The results of differential cross section for center-of-mass energies $W = 4.3$ and 4.8 GeV
are demonstrated in Fig.~\ref{fig:dcs}(a) and Fig.~\ref{fig:dcs}(b), respectively.
We noted that the angular distribution of the final $J/\psi$ for $W = 4.3$ GeV is flat, owing
to the dominance of the $s$-channel $N^*_{c\bar{c}}(4261)$ production and only S-wave involved in $N^*_{c\bar{c}}(4261)$ decays.
While the differential cross section at $W = 4.8$ GeV deviates sizably from the flat distribution, especially for backward scattering
angles, which indicates the more important role of the $u$-channel production of the nucleon pole as the largest background
for the differential cross section around $W = 4.8$ GeV. We suggest the J-PARC test all the predictions given here in future
experiments.

\section{Conclusions}
In the present work, we have made a theoretical study of the $N^*_{c\bar{c}}(4261)$ production via
$\pi^- p$ scattering. Working with an effective Lagrangian approach and resonance model,
we computed the total and differential cross sections of the $\pi^- p \to J/\psi n$ reaction
by including contributions from both $N^*_{c\bar{c}}(4261)$ and nucleon pole.
The results for the total cross section show that its crest value can exceed 11 $\mu b$ and a strong $N^*_{c\bar{c}}(4261)$ signal
appears around 4.3 GeV, which is well distinguished from the background.
For the background contribution, we found it mainly comes from the $u$-channel production of the nucleon pole.
Furthermore, it should be noted that the dependence of the differential cross section on the scattering angle for the background 
is quite different from that for the $N^*_{c\bar{c}}(4261)$ signal at backward scattering angles.

Our theoretical study here suggests that the $\pi^- p \to J/\psi n$ reaction
provides a very good place for looking for and studying the $N^*_{c\bar{c}}(4261)$ resonance.
Theoretical predictions given in this article would provide valuable information for
identifying the relevant physics of the $N^*_{c\bar{c}}(4261)$ resonance in the future experiments.
Therefore we suggest the J-PARC pay good attention to
the search for this hidden charm baryon in the $\pi^- p \to J/\psi n$ channel.
If it is observed, the information on the $N^*_{c\bar{c}}(4261)N J/\psi$ coupling may
shed some light on the nature of the $N^*_{c\bar{c}}(4261)$ resonance.

\section*{Acknowledgments}
We would like to thank Ju-Jun Xie for useful discussions.
This work is partly supported by the National Natural Science
Foundation of China under Grants Nos. 11247298, 11447105.

\bibliographystyle{unsrt}

\end{document}